\documentclass[aps,prb,a4paper,preprint,showpacs]{revtex4}
\usepackage{graphicx}
\usepackage{amsmath}
\usepackage{amsfonts}
\usepackage{amssymb}
\usepackage{amsbsy}
\usepackage{bm}

\begin{document}
\title{Tailoring the topological details of the magnetic skyrmion by the spin configuration at the edges }
\author{Yin-Yan Zhang and Rui Zhu\renewcommand{\thefootnote}{*}\footnote{Corresponding author.
Electronic address:
rzhu@scut.edu.cn}}
\affiliation{ Department of Physics, South China University of Technology,
Guangzhou 510641, People's Republic of China}

\begin{abstract}

The magnetic skyrmion structure can be formed in the chiral magnets (CMs) with strong Dzyaloshinskii-Moriya interactions. In this work, we propose a way of artificially tailoring the topological details of the skyrmion such as its radial and whirling symmetric patterns by external magnetic fields besieging the CM slab. As long as the boundary magnetic fields are strong enough to fix the boundary ferromagnetism, the attained skyrmion profile is stable over time. The dynamics of spins is considered by numerically solving the non-equilibrium Landau-Lifshitz-Gilbert equation.
\end{abstract}

\pacs {75.78.-n, 72.25.-b, 71.70.-d}

\maketitle

\section{Introduction}

The skyrmion is a topologically-protected spin texture with promising spintronic applications\cite{NagaosaNatNano2013, JonietzScience2010, ZhangSR2015}. Since the first experimental report identified the skyrmion crystal structure in the chiral magnet (CM), various materials hosting the skyrmion are found or fabricated with different skyrmion-generating mechanisms\cite{NagaosaNatNano2013, MuhlbauerScience2009, XZYuNature2010}. Multiple skyrmions can form into two-dimensional hexagonal, triangular, or square crystal structures, which are similar to the superfluid whirls forming vortex lattices in the type II superconductor\cite{MuhlbauerScience2009, SekiScience2012, MunzerPRB2010}. It is also found that the skyrmion-antiskyrmion pairs are formed in the CM as the magnetic field is increased before it stabilizes in the skyrmion crystal state\cite{KoshibaeNatComm2016}. The topological property of a spin texture can be described by the surface integral of the solid angle of the unitary spin-field vector ${\bf{n}}({\bf{r}})$ as $S = \frac{1}{{4\pi }}\int {{\bf{n}} \cdot \left( {\frac{{\partial {\bf{n}}}}{{\partial x}} \times \frac{{\partial {\bf{n}}}}{{\partial y}}} \right){d^2}{\bf{r}}} $, which is called the skyrmion number. Concerning the symmetry of the skyrmion, one can write the spin field of the skyrmion in a general form of ${\bf{n}}\left( {\bf{r}} \right) = \left[ {\cos \Phi \left( \varphi  \right)\sin \Theta \left( r \right),\sin \Phi \left( \varphi  \right)\sin \Theta \left( r \right),\cos \Theta \left( r \right)} \right]$, with $r$ and $\varphi$ the polar coordinate in the real space, $\Theta$ and $\Phi$ the polar and azimuthal angles of the local spin. The skyrmion number of a single skyrmion can be obtained as\cite{NagaosaNatNano2013}
\begin{equation}
 S = \frac{1}{{4\pi }}\left[ {\cos \Theta \left( r \right)} \right]_{r = \infty }^{r = 0}\left[ {\Phi \left( \varphi  \right)} \right]_{\varphi  = 0}^{\varphi  = 2\pi }.
 \label{skyrmion number}
\end{equation}
  While the center spin points upward and the edge spin points downward, or vice versa, different whirling patterns can be determined by the vorticity and a phase factor $\gamma $ with
\begin{equation}
  \begin{array}{*{20}{c}}
{m = {{\left. {\left[ {\cos \Theta \left( r \right)} \right]} \right|_{r = \infty }^{r = 0}} \mathord{\left/
 {\vphantom {{\left. {\left[ {\cos \Theta \left( r \right)} \right]} \right|_{r = \infty }^{r = 0}} 2}} \right.
 \kern-\nulldelimiterspace} 2},}&{\Phi \left( \varphi  \right) = \xi \varphi  + \gamma .}
\end{array}
\label{Topological Details}
\end{equation}
   Different skyrmion structures correspond to different vorticity such as $m=\pm 1$ and different helicity such as $\gamma=0,\pm\pi/2$ and $\pi$. If the in-plane component of the spin measured by $\Phi \left( \varphi  \right)$ rotates by $2 \pi \xi $ with $ \xi \ge 2$, the skyrmion acquires a high helicity and a skyrmion number $S \ge 2$. It is predicted that such a state can be dynamically created and stabilized by applying a vertical spin-polarized current\cite{XZhangPRB2016}. And reversal of the helicity of the skyrmion by thermal activation is realized in a skyrmion crystal\cite{YZYuPRB2016}. However, the multiple variants of the skyrmion defined by $\gamma$ with the same vorticity $m$ are not discovered or artificially created yet. In this letter, we will show the results of different skyrmion structures created under special edge spin configurations.

As the skyrmion cannot be continuously deformed into a homogeneous magnetic state or vice versa, the current-induced skyrmion dynamics, which imbue nontrivial topology into or drain nontrivial topology from the original magnetic states draw a great attention\cite{RommingScience2013, TchoePRB2012}. It is predicted by simulation that the skyrmion can be generated from the topologically-trivial state such as a ferromagnet and a helimagnet by external Lorentzian and radial spin currents\cite{TchoePRB2012}. The experimentally-realized skyrmion generation by the current flowing from the scanning tunneling microscope can be interpreted by the spin current extending radially from the tip point\cite{RommingScience2013}. The spatial divergent spin current can also generate the skyrmion from stripe domains\cite{WJangScience2016}. Besides spin currents, topology of the spatial distribution of the magnetic fields and constriction geometry can also generate or tailor the skyrmion out of topologically-trivial magnetic states\cite{IwasakiNatNano2013, JLiNatComm2014}. Although the skyrmion cannot be generated by continuous variation from the topologically-trivial state, it is predicted that transformation is possible between different topologically-nontrivial states such as conversion between the domain-wall pair and the skyrmion, between the skyrmion and antiskyrmion, and between the skyrmion and the bimeron\cite{YZhouNatCommun2014, ZhangSR2015}. These previous findings hint to us that topology can be seen as a tailorable  property. Direct transformation between spin textures with trivial and nontrivial topologies is forbidden\cite{RommingScience2013, IwasakiNatNano2013}. However, nontrivial topology from spin currents and domain-wall pairs can be transformed into the topological pattern of the skyrmion. And in this work, we will show that the nontrivial boundary topology can also be used to tailor the skyrmion spin texture.

The spin dynamics in the CMs can be described by the Landau-Lifshitz-Gilbert (LLG) nonlinear evolution equation including the effects of the ferromagnetic exchange, the external magnetic field, the Dzyaloshinskii-Moriya (DM) interaction, and the Gilbert damping\cite{RalphJMMM2008, EverschorPRB2012, TchoePRB2012, LakshmananPTRSA2011}. The ferromagnetic Heisenberg interaction is the origin of symmetric magnetism, with adjacent spins tending to align parallel to each other in order to decrease the exchange energy. The DM interaction in non-centrosymmetric magnets competes with the Heisenberg interaction and induces the spiral spin texture, while a perpendicular external magnetic field stabilizes the Skyrmion state with an integer topological number\cite{NagaosaNatNano2013}.
Together with theoretical simulations the artificial generation of the skyrmion have been realized in the CM and magnetic thin films by different physical mechanisms\cite{RommingScience2013, WJangScience2016, JLiNatComm2014}. However, we think that there remain two important issues less considered: 1. Present proposals require direct electric or magnetic interactions with the sample, which complicates its maintenance and possible variations; 2. Previous focus lies in whether-or-not a skyrmion is generated and differences among the many skyrmion variants are less heeded. In this work, we present a way of artificially controlling the generation of different skyrmion variants by modulating the boundary magnetic configurations. The multiple skyrmion variants bear different topological properties and have promising applications in spintronic logical gates\cite{ZhangSR2015}. Also, in this way direct interaction with the sample is avoided. Our research is motivated by the recent paper about the current-induced skyrmion dynamics in constricted geometries, in which the geometrical constriction plays an important role in forming topologically-nontrivial surroundings\cite{IwasakiNatNano2013}. We extend the subtle spatial geometry into a subtle magnetic geometry and propose the present scheme tailoring the topological details of the skyrmion. We have fabricated a two-dimensional square-lattice CM confined by patterned ferromagnetic boundaries and numerically solved the LLG equation to obtain the final stable skyrmion state.

\section{Theoretic Formalisms and Numerical Methods}

We firstly put the CM in the nonequilibrium ferromagnetic or helimagnetic state by a relatively high temperature or large magnetic fields. Then we fix the boundary ferromagnetism pattern by external magnetic fields outside of the considered CM sample and decrease the temperature or the inside magnetic field and simulate the evolution of the spin states. To experimentally realize the edge spin configuration we could fabricate Ni/Cu(001) or Fe/Cu(001) film on top of the edge of the CM\cite{JLiNatComm2014, SchulzPRB1994}. As a result of surface anisotropy and magnetoelasticity, the change between perpendicular and in-plane orientations of the easy axis of the magnetization by varying the film thickness of the magnetic metal has been reported\cite{SchulzPRB1994}. By tailoring the thickness and crystal orientation of the magnetic film covering the edge of the CM slab, the in-plane or out-of-plane magnetization can be obtained and it induces different edge spin configurations of the CM we simulated. The external magnetic field perpendicular to the CM is fixed at $0.01J \sim 0.17$ T substantial both to stabilize the skyrmion and pin the magnetization of the edge leads. The proposed scheme is shown in Fig. 1.

We consider the spin dynamics of a thin slab of CM, assuming translational symmetry along the thickness direction. The ground state is determined by the antisymmetric DM interaction, which results in the helical spin texture and can be described by the Hamiltonian\cite{IwasakiNatNano2013},
\begin{equation}
\begin{array}{l}
H =  - J\sum\limits_{\bf{r}} {{{\bf{M}}_{\bf{r}}} \cdot \left( {{{\bf{M}}_{{\bf{r}} + {{\bf{e}}_x}}} + {{\bf{M}}_{{\bf{r}} + {{\bf{e}}_y}}}} \right)} \\
 - D\sum\limits_{\bf{r}} {\left( {{{\bf{M}}_{\bf{r}}} \times {{\bf{M}}_{{\bf{r}} + {{\bf{e}}_x}}} \cdot {{\bf{e}}_x} + {{\bf{M}}_{\bf{r}}} \times {{\bf{M}}_{{\bf{r}} + {{\bf{e}}_y}}} \cdot {{\bf{e}}_y}} \right)} \\
 - {\bf{B}} \cdot \sum\limits_{\bf{r}} {{{\bf{M}}_{\bf{r}}}} .
\end{array}
\end{equation}
The local magnetic moment $\bf{M_{\bf{r}}}$ is defined as $\bf{{M_r}} \equiv  - \bf{{S_r}}/\hbar$ (where $\bf{{S}_{r}}$ is the local spin at $\bf{r}$). It can be treated as classical vectors whose length is fixed as ${\bf{|{M_r}|}}= M$. Therefore ${{\bf{M}}_{\bf{r}}} = M{\bf{n}}\left( {\bf{r}} \right)$ with ${\bf{n}}\left( {\bf{r}} \right)$ the unitary spin field vector. $J$ and $D$ correspond to the ferromagnetic and DM exchange energies, respectively. $\bf{B}$ is the strength of the external magnetic field contributing to the Zeeman term. ${\bf{e}}_x$ and ${\bf{e}}_y$ are the unitary vectors in the $x$ and $y$ directions within the CM slab plane, respectively. The magnetic dynamics of the considered model can be described by the LLG equation\cite{IwasakiNatNano2013},
\begin{equation}
\frac{{d{{\bf{M}}_{\bf{r}}}}}{{dt}} =  - \gamma {{\bf{M}}_{\bf{r}}} \times {\bf{B}}_{\bf{r}}^{{\rm{eff}}} + \frac{\alpha }{M}{{\bf{M}}_{\bf{r}}} \times \frac{{d{{\bf{M}}_{\bf{r}}}}}{{dt}},
\label{LLG Equation}
\end{equation}
which is formulated in the continuum approximation. Here $\gamma $ is the gyromagnetic ratio, ${{\bf{B}}_{\bf{r}}^{{\rm{eff}}}}$ represents the effective field defined by ${\bf{B}}_{\bf{r}}^{{\rm{eff}}} =  - \frac{1}{{\hbar \gamma }}\frac{{\partial H}}{{\partial {{\bf{M}}_{\bf{r}}}}}$, and $\alpha$ is the Gilbert damping parameter\cite{RalphJMMM2008, HalsPRB2014}. We choose the unit of time $t_0 =\hbar/J$. Evolution of the spin field is solved by the forth Runge-Kutta method.

\section{Numerical Results and Interpretations}

Numerical results of the simulation under different boundary and initial conditions are given in this section. The final states shown in Fig. 2 correspond to the ferromagnetic boundaries with the spin pointing upward or downward perpendicular to the CM slab. We adopt a square lattice of size $30 \times 30$ in the unit of the lattice constant throughout the simulation. We set the initial state of the CM to be a helimagnet with spiraling spin field ${\bf{n}}\left( {\bf{r}} \right) = \left[ {\cos \left( {Qx} \right),\sin \left( {Qx} \right),0} \right]$ in the Cartesian coordinate with $Q=2 \pi /17.8$ to sustain a complete spiral within the slab. We found that with the boundary magnetism fixed the value of $Q$ only slightly affects the evolution process and the time needed to achieve the final state without altering the final stable state. The typical parameters $D=0.18 J$, $\alpha=0.01$ and ${\bf{B}}=B {\bf {e}}_z$ with $B=0.01 J$ are used. The $z$-component magnetic field is within the range of the skyrmion phase in a CM with other parameters fixed\cite{IwasakiNatNano2013}. Without a noticeable error accumulation, we found the helical state collapsed rapidly as a result of reflection by the fixed boundary. The spins adjacent to the boundary are forced to follow the direction of the boundary spins, which are fixed by top ferromagnetic leads surrounding the CM slab. Then the influence of the boundary arrives at the center of the CM in a relay and fluctuating process until an isolated skyrmion or antiskyrmion is created, whose skyrmion number is very closed to 1 or -1. It can be seen that the skyrmion in (a) whirls counterclockwisely corresponding to $m=1$ and $\gamma =\pi /2$ and the antiskyrmion in (b) whirls clockwisely corresponding to $m=-1$ and $\gamma =- \pi /2$. The whirling pattern is determined by the direction of the DM interaction. When $D>0$, the $S=1$ state is always a counterclockwise vortex and vice versa. These two states are already confirmed by experiment\cite{NagaosaNatNano2013, MuhlbauerScience2009, XZYuNature2010}.

To show that the obtained skyrmion state is stabilized by the external magnetic field, variation of the skyrmion number in time is shown in Fig. 3. As the initial state and other parameters are the same, we attribute the topological type of the skyrmion generated to the applied ferromagnetic boundary condition. The ferromagnetic boundary can break the original trivial topology. Nontrivial topology emerges as the spins scatter with the boundary fixed spins. However, it can be seen from Fig. 1 that the boundary is tailored into a hard circle topologically nontrivial. It can also be seen in Fig. 3 that even when the center region is a helimagnet or a ferromagnet in the initial state of $t=0$, the boundary contributes a finite Chern number although it is not an integer. This is also the case in the domain-wall pair reversibly transforming into a skyrmion\cite{YZhouNatCommun2014}. The nontrivial topology with partial Chern number is not stable and evolves into a full skyrmion in a short period of back-and-forth interactions among spins, which can be observed in the evolution of $S$. As the strength of the DM interaction $D=0.18 J$ is used, the diameter of the naturally-host skyrmion in a CM is $\lambda  =D/J \approx 35$ in the unit of $a$, which strengthens the skyrmion and antiskyrmion obtained  with the same radius of $15$.

Besides the usual skyrmion and antiskyrmion spin profiles, our simulations show some variants of the skyrmion with the vorticity and helicity defined by special $\gamma $ in Eq. (\ref{Topological Details}), which exactly match the spin profile proposed earlier\cite{NagaosaNatNano2013, BogdanovJMMM1999}. The results are shown in Fig. 4. We firstly apply a strong magnetic field to form a ferromagnetic initial state with the local magnetic moments pointing in the $z$-direction as shown in Fig. 1(b). Then we abruptly turn down the magnetic field to $B=0.01 J$ and take the ferromagnetic state as the initial state and apply the boundary magnetic leads to induce the ferromagnetic boundaries and tailor the topological details of the skyrmion state. One kind of the edge spin configuration with the edge spins lying in-plane and winding counterclockwisely is shown in Fig. 1(b), which corresponds to the final state of Fig. 4 (d). The other types of ferromagnetic boundaries we considered are that the boundary spins all lie in-plane pointing out of the center, toward the center, and winding clockwisely, which correspond to the final states of Fig. 4(a) (b), and (c), respectively. For these four cases, $D=0.01 J$, which is smaller than the cases shown in Fig. 2. It is because that the boundary-determined final state is a vortex with $S=\pm 1/2$. We can see from Eq. (\ref{skyrmion number}) that when the spin at the edge of the skyrmion(vortex) lies in-plane instead of pointing upward or downward, $\cos \Theta \left( \infty  \right) = \cos \left( { \pm {\pi  \mathord{\left/
 {\vphantom {\pi  2}} \right.
 \kern-\nulldelimiterspace} 2}} \right) = 0$, contributing an $m=\pm 1/2$ in Eq. (\ref{Topological Details}). The state is a magnetic vortex. As the spiraling period in the CM is determined by $D/J$, in the considered case, the radius of the full skyrmion extends much farther than the $\Theta  = 0$ edge giving rise to much longer spiraling period. Therefore, only a much smaller $D/J$ can sustain the longer spiraling period. In real materials, $D/J$ varies between a large range, which promises the workability of our proposition. As shown in Fig. 4, under interference with the robust boundary, the initial ferromagnetic states have broken their original trivial topological structures and evolved into the magnetic vortex state with $S$ close to $0.5$. The spins adjacent to the boundary adiabatically follow the fixed boundary direction and the Heisenberg and DM interactions conduct their effects in a relay pattern generating a vortex spin structure. Using the boundary magnetism, the topological details of the skyrmion or vortex structure are precisely tailored. The stable magnetic vortex states are obtained with $m=1/2$ and $\gamma =0$, $\pi$, $- \pi /2$, and $\pi /2$, respectively, which are shown in Fig. 4 (a) to (d). The value of $\gamma$ can be controlled by artificial ferromagnetic boundaries.

 Again, to confirm the nonequilibrium stableness of the magnetic vortex obtained, the time evolution of the skyrmion number is shown in Fig. 5. It can be seen from the skyrmion number that the evolution of the four kinds of spin states are quite similar. Without the ferromagnetic boundary, the spins were supposed to form an antivortex with negative skyrmion number under the emergent electromagnetic field. When subject to the fixed boundary, the spins receive a great influence and rapidly swing into a vortex, hinting that the artificial effect of the external magnetic boundaries is strong. The topological details of different spin structures can be tailored in this way.

\section{Conclusions}

In this work, we have investigated the spin dynamics in the CM confined by different types of ferromagnetic boundaries. Particularly, by applying the ferromagnetic boundaries with the boundary spins pointing downward; upward; out of the center; toward the center; winding clockwisely; and counterclockwisely, we have obtained the skyrmion, antiskyrmion, and vortex structures with the vorticity and helicity $m=1$, $\gamma =\pi /2$; $m=-1$, $\gamma =- \pi /2$; $m=1/2$, $\gamma =0$; $\pi$; $-\pi /2$; $\pi /2$. By manipulating the ferromagnetic boundary, we can artificially control the generation of different types of the skyrmion spin structures, which provides a new way to produce and tailor the skyrmions in the CM.

\section{Acknowledgements}

This project was supported by the National Natural Science
Foundation of China (No. 11004063) and the Fundamental Research
Funds for the Central Universities, SCUT (No. 2014ZG0044).

\clearpage

\begin{figure}[ht]
\includegraphics[height=6cm, width=10cm]{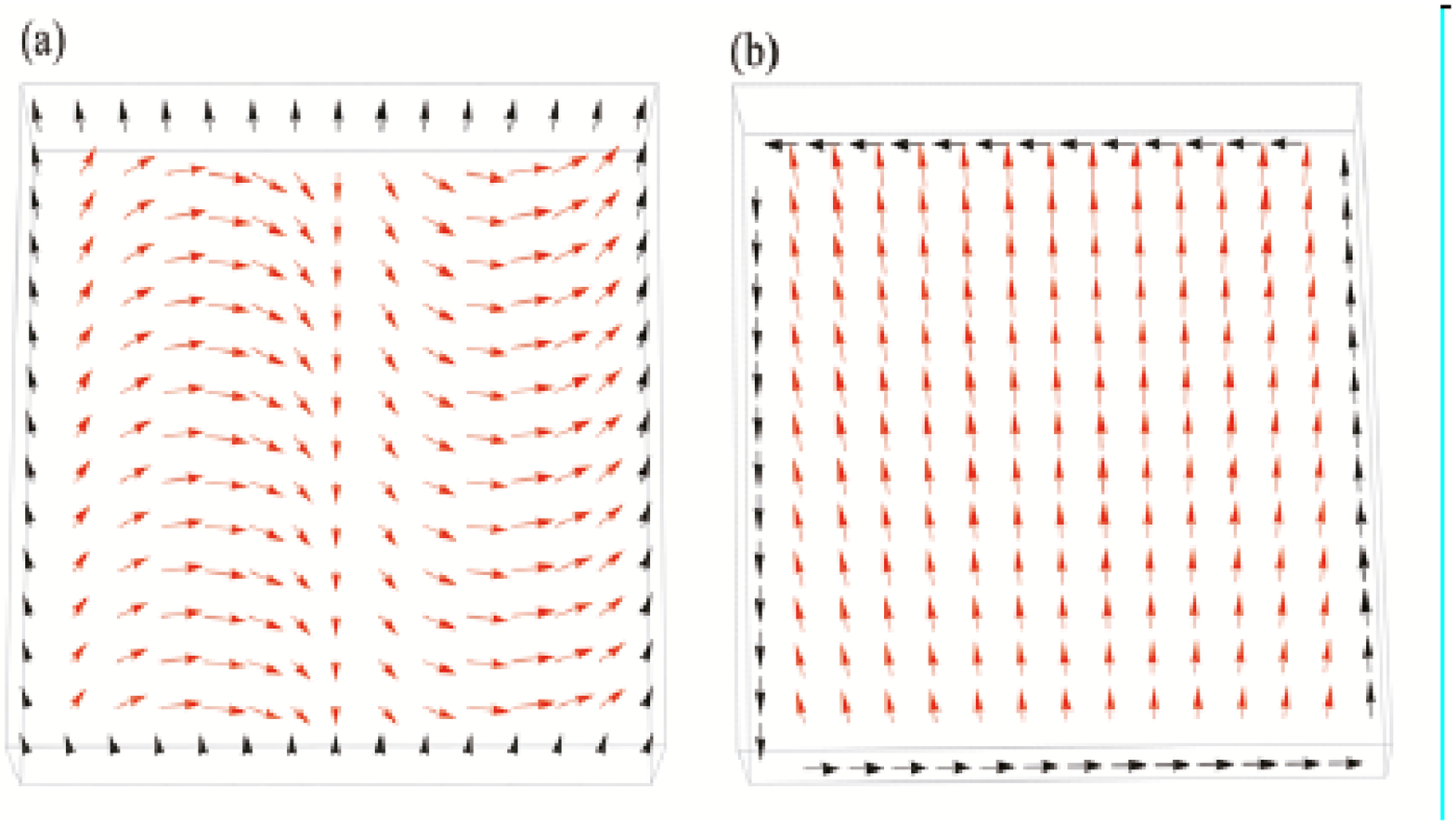}
\caption{ Sketch of the CM confined by different ferromagnetic boundary patterns with (a) helimagnetic and (b) ferromagnetic initial states. The black arrows correspond to boundary ferromagnetic patterns and the red arrows correspond to the initial spin states of the CM. In panel (a), all the surrounding spins point upward. In panel (b), The surrounding spins point within the CM plane and wind counterclockwisely. The spin configuration at the edges in both of the two panels imbue nontrivial topology into the middle CM slab. }
\end{figure}

\begin{figure}[ht]
\includegraphics[height=10cm, width=12cm]{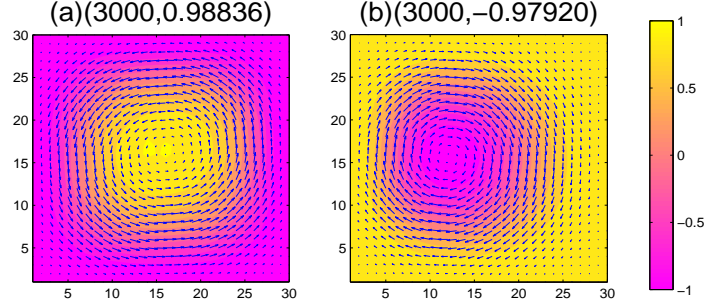}
\caption{ Final states of the CM confined by (a) downward and (b) upward ferromagnetic boundaries. (a) corresponds to a skyrmion with $m=1$ and $\gamma =\pi /2$ and (b) corresponds to an antiskyrmion with $m=-1$ and $\gamma=-\pi /2$. $m$ and $\gamma$ are defined in Eq. (\ref{Topological Details}). The time $t$ in the unit of $t_0$ together with the skyrmion number $S$ in the form of $(t,S)$ is labeled upper to each panel. The color distribution indicates magnitude of the z-component of the spins. The initial state of the CM are a helimagnet in both panel (a) and (b). }
\end{figure}

\begin{figure}[ht]
\includegraphics[height=8cm, width=12cm]{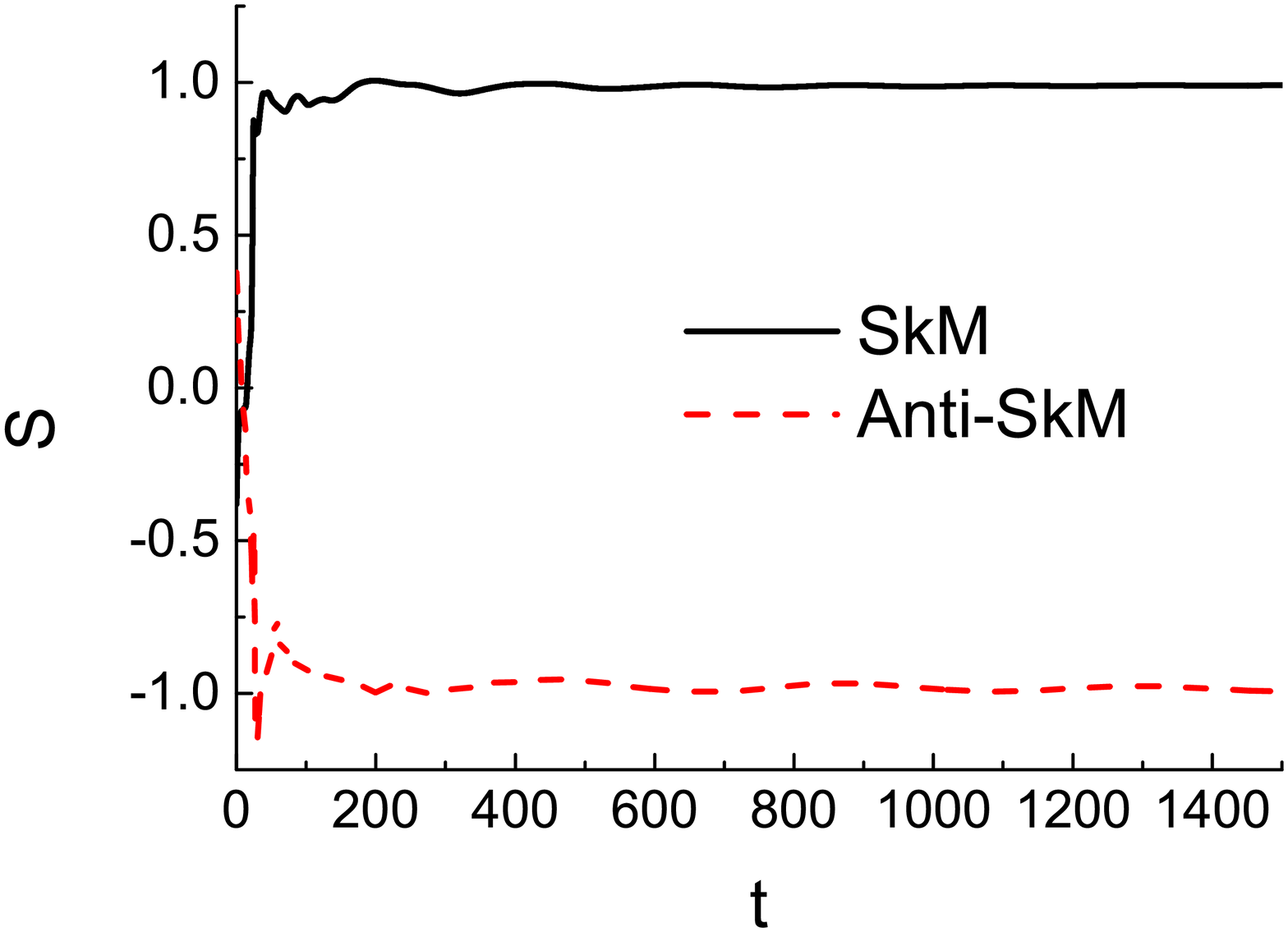}
\caption{ Time dependence of the skyrmion number $S$ for the two cases considered in Fig. 2. After the early stage of oscillation, the spins enter into a stable state whose skyrmion number stays at $+1$ or $-1$ with the edge spin configuration fixed.}
\end{figure}

\begin{figure}[ht]
\includegraphics[height=10cm, width=12cm]{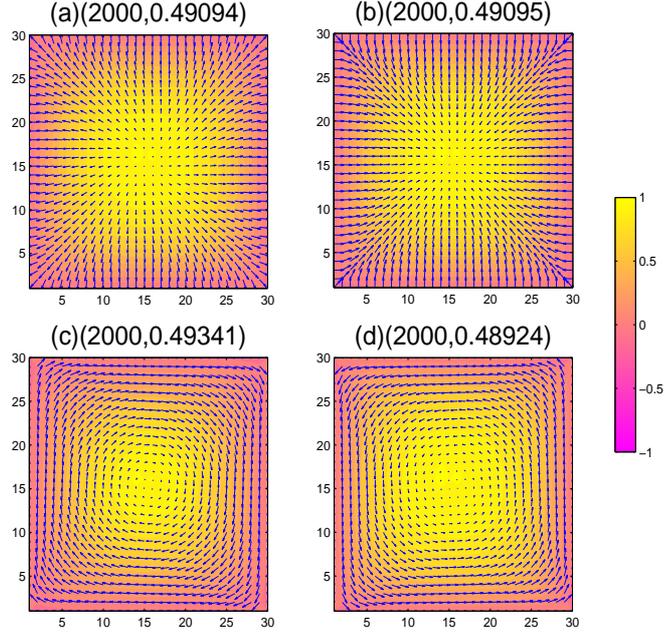}
\caption{ The final states of the magnetic vortex structures corresponding to in-plane ferromagnetic boundaries with the boundary spins (a) pointing out of the center, (b) toward the center, (c) winding clockwisely, (d) winding counterclockwisely. The final states shown in panels (a) to (d) correspond to the vorticity $m=1/2$ and the helicity $\gamma=0$, $\pi$, $-\pi/2$, and $\pi/2$ defined in Eq. (\ref{Topological Details}), respectively. }
\end{figure}

\begin{figure}[ht]
\includegraphics[height=8cm, width=14cm]{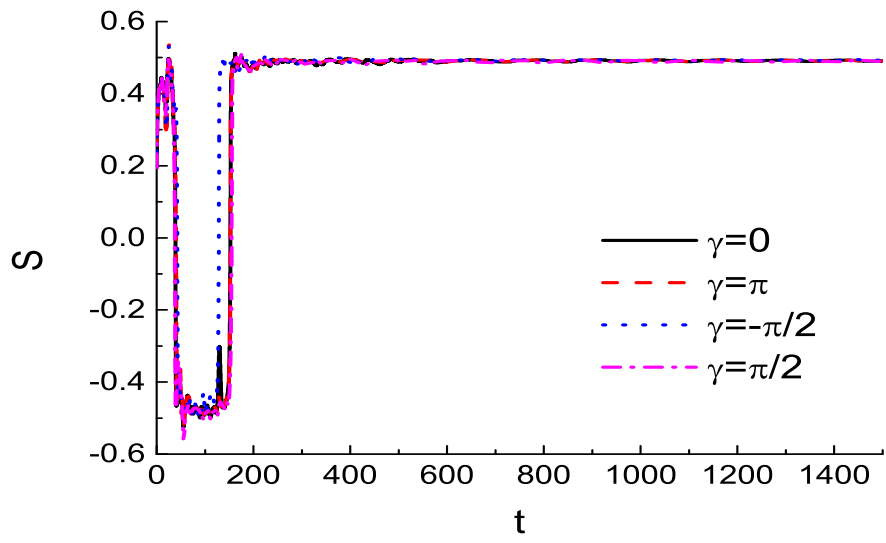}
\caption{ Time dependence of the skyrmion number $S$ of the four kinds of in-plane ferromagnetic boundaries corresponding to Fig. 4. }
\end{figure}

\end{document}